\documentclass[12pt]{article}   
\usepackage{epsfig}   
\newcommand{\eq}{\begin{equation}}   
\newcommand{\eqx}{\end{equation}}   
\newcommand{\eqn}{\begin{eqnarray}}   
\newcommand{\eqnx}{\end{eqnarray}}

\title{{\bf QCD at Photon Colliders} \footnote{Introductory talk given at 
the International Workshop on High Energy Photon Colliders, 14th - 17th 
June 2000,  
DESY, Hamburg, Germany}}   
\author{{\sc J.~Kwieci\'nski} \footnote{e-mail: jkwiecin@solaris.ifj.edu.pl}    
        \\   
        \it Department of Theoretical Physics,\\   
                 \it H.~Niewodnicza\'nski Institute of Nuclear Physics,\\   
                 \it 31-342 Cracow, Poland}   
\date{}   
\begin{document}   
\maketitle   
\begin{abstract}   
The novel possibilities of probing the   photon structure 
and high energy limit of QCD at photon colliders are summarised.  
We discuss  the photon structure function $F_2^{\gamma}(x,Q^2)$, 
the gluon distribution in the photon and the spin dependent structure function 
$g_1^{\gamma}(x,Q^2)$ of the photon and emphasise advantages of the 
photon colliders for measuring these quantities.     
The possibility of probing 
the 
QCD pomeron and odderon in $\gamma\gamma$ and $e \gamma$ collisions 
is also described. 
\end{abstract}   
\section{Introduction}
The photon colliders offer unique possibility to probe QCD 
in a new and  hitherto unexplored  regime.  
Very high luminosity and relatively sharp spectrum of the 
backscattered laser photons are the enormous advantages 
of the photon colliders making it possible to measure very
precisely various quantities as well as to have an access to
rare QCD processes. The aim of this 
introductory talk is to elaborate the following two broad
subjects for which the possible measurements at 
photon colliders may be  particularily relevant:  
\begin{enumerate}
\item The photon structure. 
\item High energy QCD. 
\end{enumerate}
The content of our talk is as follows: in the next  section 
we discuss the deep inelastic $e\gamma$ scattering as the tool 
for probing the quark content of the photon while in Section 3 
we consider the measurements which 
may probe the gluon distributions in the photon.   
In Section 4 we very briefly summarise the theoretical
expectations concerning the high energy behaviour of QCD 
(i.e. the QCD pomeron and odderon) and  discuss the 
possibilities of the photon colliders to probe these 
expectations. We shall in particular consider  production of vector mesons  in $\gamma \gamma$ and 
$\gamma^* \gamma$ collisions  and  forward jet measurement in $\gamma^* \gamma$ collisions.   
as a probe of the QCD pomeron. 
In Section 5 we discuss the spin dependent 
 structure function $g_1^{\gamma}(x,Q^2)$ of the photon .  
Finally in Section 6  we present a brief 
summary of our talk. 
   
\section{Deep inelastic $e\gamma$ scattering} 






The 
deep inelastic $e~\gamma $ scattering, i.e. the process: 
$$e(p_e) ~ + ~\gamma (p_{\gamma}) \rightarrow e(p_e^{\prime}) + X$$
is characterised by the following kinematical variables: 

$$ 
s=(p_e + p_{\gamma})^2,~~~~ q=p_e - p_e^{\prime},~~~~ Q^2=-q^2,~~~~ 
W^2=(q+p_{\gamma})^2
$$

\begin{equation}
y={p_{\gamma}q\over p_ep_{\gamma}},~~~~x={Q^2\over 2p_{\gamma}q}
\label{diskin}
\end{equation}

  Possibility of precise measurement of the kinematical variables $x,Q^2$ 
in $e\gamma$ 
DIS is an enormous advantage of  photon colliders. It is 
linked with the form of the spectrum of the backscattered laser 
photons which is more advantageus than the Weizs\"aker - 
Williams spectrum  of the (quasi) real photons exchanged   
in single tagged $e^+e^-$ collisions. 
This in particular allows   precise measurement of the 
photon structure function(s) with much better accuracy than 
in the single tagged  $e^+e^-$ collisions.  \\

The deep inelastic $e\gamma$ scattering is related in the
following way 
to the photon structure functions $F_2^{\gamma}(x,Q^2)$ and $F_L^{\gamma}(x,Q^2)$:
  
$$
{d\sigma(e\gamma \rightarrow eX) \over dE_{tag} dcos\theta_{tag} 
}=$$

\begin{equation}
  {4\pi^2 E_{tag}\over Q^4 y} \left[\left (1+(1-y)^2\right)F_2^{\gamma}(x,Q^2)
  -y^2F_L^{\gamma}(x,Q^2) \right]
\label{discx}
\end{equation}
where $E_{tag}$ and   $\theta_{tag}$ denote the energy and the
scattering angle 
of the tagged electron. \\
 
In the leading logarithmic approximation (LLA) of perturbative QCD the
structure 
function $F_2^{\gamma}(x,Q^2)$ is derictly related to the quark 
 distributions in the photon:

\begin{equation}
F_2^{\gamma}(x,Q^2)=2x\sum_ie_i^2q_i^{\gamma}(x,Q^2) 
\label{f2}
\end{equation}

where $e_i$ are the quark charges.   
Evolution of parton densities  is governed by the
Altarelli-Parisi equations which in  the leading order take the following 
form \cite{WITTEN,ZERWAS,LAC,YELLOWGG}:    

$$
Q^2{dq_i^{\gamma}(x,Q^2)\over dQ^2} = 
{\alpha_{em} N_c\over 2\pi} e_i^2 k_q^{\gamma}(x) + 
{\alpha_s(Q^2)\over 2\pi} 
\left(P_{qq} \otimes q_i^{\gamma} + P_{qg} 
\otimes g^{\gamma}\right) 
$$

\begin{equation}
Q^2{dg^{\gamma}(x,Q^2)\over dQ^2} =  
{\alpha_s(Q^2)\over 2\pi} 
\left(P_{gq} \otimes \sum_i
q_i^{\gamma}
  + P_{gg} \otimes g^{\gamma} \right)
\label{ap}
\end{equation}

where 

\begin{equation}
k_q^{\gamma}(x) = x^2 + (1-x)^2
\label{kqx}
\end{equation}

and 

\begin{equation}
P_{ij} \otimes f =\int_x^1 {dz\over
z}P_{ij}\left(z\right)f\left({x\over z}
,Q^2\right)
\label{convol}
\end{equation}
The evolution equations (\ref{ap}) involve, besides the quark
 distributions  also the gluon distribution $g^{\gamma}(x,Q^2)$ in the photon.\\

The novel feature of the evolution equations for the parton
distributions in the photons is the presence of the  
inhomogeneous term proportional to $k_q^{\gamma}(x)$ which corresponds to   
the point-like coupling 
of the photon to quarks (antiquarks).  The solution of equations
(\ref{ap}) for the parton distributions $f^{\gamma}(x,Q^2)$ 
 in the photon 
($f^{\gamma}=q_i^{\gamma},g^{\gamma}$) can  be written in the following form:

\begin{equation}
f^{\gamma}(x,Q^2)=f_{pl}^{\gamma }(x,Q^2) + f_{had}^{\gamma}(x,Q^2)
\label{sol}
\end{equation}
 
In this equation the distributions $f_{pl}^{\gamma }(x,Q^2)$ 
are exactly calculable 
(modulo dependence upon the 
scale $Q_0^2$ where $f_{pl}^{\gamma}(x,Q_0^2)=0$ ). The hadronic component 
$f_{had}^{\gamma}(x,Q^2)$, which corresponds to the general
solution of the homogeneous 
evolution equatioms depends upon the non-calculable starting distributions 
 at the input scale $Q_0^2$. Existing QCD analyses use NLO approximation in $\bar MS$ or $DIS_{\gamma}$ 
scheme(s) \cite{GRV,GRS}. \\

In Fig. 1 we show  acceptance regions and event rates which are 
expected to be reached at $e \gamma$ DIS for the 
 backscattered-laser $e\gamma$ mode of a 500 GeV linear collider \cite{PC}. 
 It can be seen from this Figure that the  $e\gamma$ collider can offer  
 opportunity to probe 
 the very 
low values of $x$ ($ x\sim 10^{-4}$ for reasonably large values of 
$Q^2 \sim 10 GeV^2$). At NLO low $x$ behaviour of parton
distributions in the 
photon is dominated by a hadronic component.  On the other hand
at large values of $x$  
the quark distributions are dominated by the point-like term.  \\
At low $x$ one  has also to include the low $x$ resummation
effects in the spltting and coefficient functions which are
generated by the small $x$ (i.e. BFKL) dynamics (see Sec. 5) \cite{BVSMX}.
These small $x$ effects can significantly enhance the magnitude  of the parton 
distributions in the photon and the photon structure function(s)
at low values of $x$.   The hadronic component remains to be the
dominant contribution in this region yet the magnitude of the    
  point like contribution is also enhanced.\\

At very large values of $Q^2$   the deep inelastic $e\gamma$
scattering can also acquire significant contribution from the
$Z_0$ exchange.  It modifies formula (\ref{discx}) which
corresponds 
to the (virtual) photon exchange.      Moreover at very large values 
of $Q^2$  one can have access to the charge current effects in
the 
deep inelastic process $e ~ + ~\gamma  \rightarrow  \nu~ +~ X$ 
which is mediated by the $W$ exchange.  The study of this 
process can in particular give information about the flavour 
decomposition of the quark distributions in the photon \cite{ZERDR}.

\section{Gluon distribution in the photon}

The gluon distribution in the photon is a poorly known
quantity.  
In principle it can be determined from the analysis of the
scaling 
violation of $F_2^{\gamma}(x,Q^2)$ generated by the evolution
equations (\ref{ap}), 
but this constraint
turns out to be rather weak.  In order to obtain 
the gluon distribution it is therefore much better to study the
 measurements which are predominantly sensitive to 
the gluon content of the photon. Thus  at low $x$ the measurement of
the 
longitudinal structure function $F_L^{\gamma}(x,Q^2)$ might
provide a  ueful constraint on the gluon distributions 
in the photon.  The complementary and presumably much more 
feasible are the dedicated  measurements of the hadronic 
final states in  $\gamma\gamma$ 
collisions.  The following two processes are of particular
interest:   
 \begin{enumerate}

\item Dijet production, which contains  the process \\

$$
\gamma ~ g \rightarrow q ~ \bar q 
$$

\item  Charm production, which is sensitive to the mechanism\\

$$
\gamma ~ g \rightarrow c ~ \bar c
$$
\end{enumerate}

Since both these  processes are, at least in certain kinematical
regions dominated by the photon - gluon fusion  mechanism as indicated 
above,  
they are sensitive to the gluon distribution in the photon. 
The detailed discussion of these processes is presented in this 
Workshop  
in the talks by Thorsten Wengler and Albert De Roeck \cite{WENGLER,ALBERT}.  
\section{QCD at high energies} 
Important class of hard processes are the so called semi-hard 
processes in which the (large) scale $Q^2$ characterising the "hardness" 
is much smaller than the very large energy squared  $W^2$, i.e. 
$W^2/Q^2>>1$.  These processes are sensitive to the resummation 
of the powers of the new class of large logarithm $ln(W^2/Q^2)$
in perturbative 
QCD calculations, 
i.e. they do probe the high energy limit of perturbative QCD. \\
         
The high energy limit of perturbative QCD is at present fairly
well understood.  The dominant contribution is given by the QCD 
pomeron singularity which is generated by the ladder diagrams 
with the (reggeised) gluon exchange along the ladder \cite{BFKL}.  
Two gluon exchange gives the energy independent
cross-sections while exchange of the gluon ladder with 
interacting gluons generates  increase of cross-sections 
with energy. \\

The sum of ladder diagrams 
 generates the  Balitzkij, Fadin, Kuraev, Lipatov (BFKL)
equation:
\begin{equation}
f(x,\hat k^2) = f^0(x,\hat k^2) + {3 \alpha_s\over 2 \pi} K\otimes f 
\end{equation}
where 
$f(x,\hat k^2) $ denotes the unintegrated gluon distribution
with $x$ and $k^2$ 
denoting the longitudinal momentum fraction of the parent photon
carried by the gluon and $k^2$ is the transverse momentum
squared of the gluon. 
$f^0(x,\hat k^2)$ is the suitably defined inhomogeneous term. 
\\

In leading order (i.e. in the leading $ln(1/x)$ approximation) we have: 

\begin{equation} 
K\otimes f = \int_x^1 {dz\over z}\int {d^2\hat q\over  \pi \hat q^2} 
[f(z,(\hat k + \hat q)^2) - \Theta(\hat k^2-\hat q^2)f(z,\hat k)]
\end{equation}
 
The following properties of the BFKL dynamics embodied in the
BFKL 
equation should be emphasised. 
\begin{description}
\item{1.}Diffusion of transverse momentum along the chain 
 which should reflect itself in the hadronic final state\\
 
\item{2.} Characteristic rise with decreasing $x$. 
In LO $f \sim x^{-\lambda}, \lambda = 4ln(2) 3\alpha_s/\pi$\\
 
\item{3.} Large subleading effects \cite{BFKLNL}.  Their major part is however
understood and is 
under control \cite{RESUM}.\\
\end{description}
Besides the pomeron  QCD does also predict existence of 
the so called QCD  odderon \cite{ODDERON}.   The QCD odderon corresponds 
to the  (interacting) three gluons exchange with the three gluons 
coupled to the 
odd C, i.e. to non-vacuum quantum numbers.  The intercept of the
odderon is close to $1$. The
cross-sections generated by the odderon exchange should then be
approximately independent of the incident energy.\\

The high energy photon colliders offer  interesting  opportunities to
probe the QCD pomeron.   It may also appear possible that the 
photon colliders, thanks to their very
high luminosity will allow   to measure rare processes generated by the 
odderon exchange.\\

 The following processes  are  very useful tools for probing
the QCD pomeron in high energy $e \gamma$ and $\gamma 
\gamma$ collisions \cite{CONTRDR,GGVVNS,KMJPSI}:    
\begin{enumerate}
\item Deep inelastic   $e \gamma$ scattering accompanied by a
jet moving close to the direction of the photon ($x_j >>x$,
where $x_j$ is the momentum fraction of the parent photon
carried by a jet) and with 
$k_{Tj}^2   \sim Q^2$ \cite{CONTRDR}.\\

\item  Diffractive production of Vector Mesons in $e\gamma$ DIS,
i.e.\\
$\gamma^*  ~ \gamma \rightarrow V ~ V$\\
$\gamma^*  ~ \gamma \rightarrow V ~ X$\\

\item Diffractive production of heavy vector mesons in $\gamma ~\gamma$ 
collisions\\
$\gamma  ~ \gamma \rightarrow V ~ V $\\
$\gamma  ~ \gamma \rightarrow V ~ X$\\
\end{enumerate}

In Fig. 2 we show  
theoretical expectations for the  cross-section of the  diffractive 
$J/\psi $ production in $\gamma \gamma$ collisions, i.e. of the 
process $\gamma \gamma \rightarrow J/\psi J/\psi$ \cite{KMJPSI}.  
Expected magnitude of 
this cross-section suggests that the   measurement of the 
reaction $\gamma \gamma \rightarrow J/\psi J/\psi$ 
at the photon colliders 
should certainly be feasible. 
 
The following processes might probe the QCD odderon \cite{GGPPNS,KMETAC}: 

\begin{enumerate} 
\item Quasidiffractive production of pseudoscalar mesons in $e\gamma$ DIS.\\

$\gamma^*  ~ \gamma \rightarrow P ~ P$\\
$~\gamma^*  ~ \gamma \rightarrow P ~ X$\\


%
 
\item Quasidiffractive production of (heavy) pseudoscalar mesons in $\gamma ~\gamma$ 
collisions\\

$\gamma  ~ \gamma \rightarrow P ~ P$\\

$\gamma  ~ \gamma \rightarrow P ~ X$\\

\end{enumerate}

\section{Spin dependent structure function $g_1^{\gamma}(x,Q^2)$}

Thanks to the possibiliity of having polarised beams the photon 
colliders offer an opportunity to measure the spin dependent
structure 
function $g_1^{\gamma}(x,Q^2)$ of the photon \cite{GG1,GG2}. 
This is
completely unknown quantity 
and its measurement in polarised $e\gamma$ DIS  would be extremely useful for testing 
QCD predictions in the broad region of $x$ and $Q^2$.\\
 
The spin dependent parton distributions in 
the photon $\Delta f^{\gamma}(x,Q^2)$ satisfy in LO 
Altarelli - Parisi equations similar 
to equations (\ref{ap}) with the  term $k_q^{\gamma}(x)$ 
and the splitting functions $P_{ij}(z)$ being replaced by their 
spin dependent counterparts  $\Delta k_q^{\gamma}(x)$ and  $\Delta P_{ij}(z)$.  
The  NLO  formalism is also available.  
In Fig. 3 we show  theoretical expectations for $\Delta u^{\gamma}(x,Q^2)$, 
$\Delta g^{\gamma}(x,Q^2)$ and for $g_1^{\gamma}(x,Q^2)$ \cite{GG2}. 
We show results for the two model assumptions concerning the input 
hadronic component $\Delta f^{\gamma}_{had}(x,Q_0^2)$ at the 
reference 
scale $Q_0^2$: 

\begin{equation}
\label{eq:maxsat} 
\Delta f^{\gamma}_{had}(x,Q_0^2) = f_{had}^{\gamma}(x,Q_0^2)\;\;,
\end{equation}
where $f_{had}^{\gamma}(x,Q_0^2)$ denotes hadronic 
component of spin independent parton distributions, and for 
\begin{equation}
\label{eq:minsat}
\Delta f^{\gamma}_{had}(x,Q_0^2) = 0
\end{equation}
 It should be observed that the high energy photon colliders would 
  make it
possible   to probe the spin dependent structure 
function of the photon $g_1^{\gamma}(x,Q^2)$ for the very small
values 
of $x$ where it 
is sensitive to the novel effects of the double $ln^2(1/x)$
resummation \cite{JKBZGG1}.  Study of some dedicated measurements, like dijet 
or charm production in polarised $\gamma \gamma$ scattering
might also give access to the spin dependent 
gluon distribution $\Delta g^{\gamma}(x,Q^2)$ in the photon.

\section{Summary and conclusions}

In this talk we have confined ourselves to a brief survey 
of the possible measurements at photon colliders 
which could probe photon structure 
and high energy limit of perturbative QCD.   
We have in particular  emphasised the following points: 
\begin{enumerate}
\item Photon colliders permit precise determination  of kinematical 
variables in $\gamma ~\gamma$ or $e~ \gamma$ scattering that
will allow  
precise measurement of quantities relevant for understanding the structure 
of the photon ($F_2^{\gamma}(x,Q^2), ~~g^{\gamma}(x,Q^2)$).  
\item Photon colliders  will give access  to the (very) small
$x$ region 
in $e\gamma$ DIS.    
\item Photon colliders  will be a unique probe of the high energy limit of 
QCD  testing effects of the QCD pomeron and possibly 
also the odderon exchanges.  The diffractive production of $J/\Psi$ 
mesons in $\gamma \gamma$ collisions is one of the very promising 
processes for probing the QCD pomeron at photon colliders.    
\item Photon colliders offer a unique possibility to measure the spin 
dependent structure function   $g_1^{\gamma}(x,Q^2)$
of the photon.    
\end{enumerate}
\section*{Acknowledgments} 
I thank Valery Telnov for his kind invitation to the Workshop. I would like 
to 
congratulate him and other organisers for preparing an excellent 
meeting. I thank Ilya Ginzburg, Maria Krawczyk, Leszek Motyka,  Albert De Roeck, 
Marco Stratmann, Thorsten Wengler, Andreas Vogt and Peter Zerwas for 
useful discussions.  
This research has been supported in part by the European Community grant   
'Training and Mobility of Researchers', Network 'Quantum  
Chromodynamics and the Deep Structure of Elementary Particles'  
FMRX-CT98-0194.

\newpage
 \begin{figure}[t]
\label{avf4}
\vspace*{-1mm}
\centerline{\epsfig{file=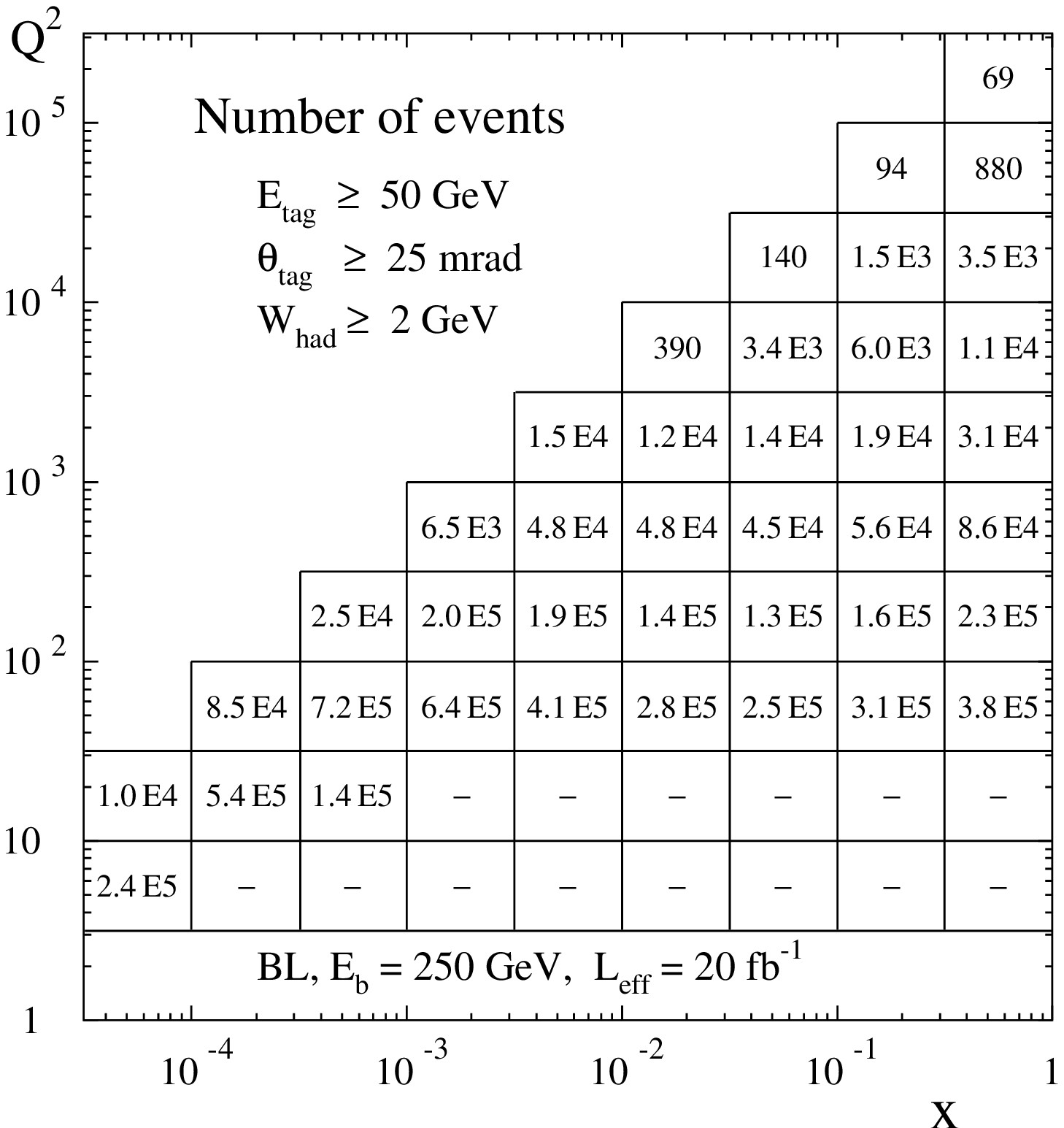,width=7.5cm,angle=0}}
\vspace*{-8mm}
\caption{}
\end{figure}
\newpage
\begin{figure}[hbpt]
\leavevmode
\begin{center}
\parbox{8cm}{
\epsfxsize = 7.5cm
\epsfysize = 7.5cm
\epsfbox[18 200 565 755]{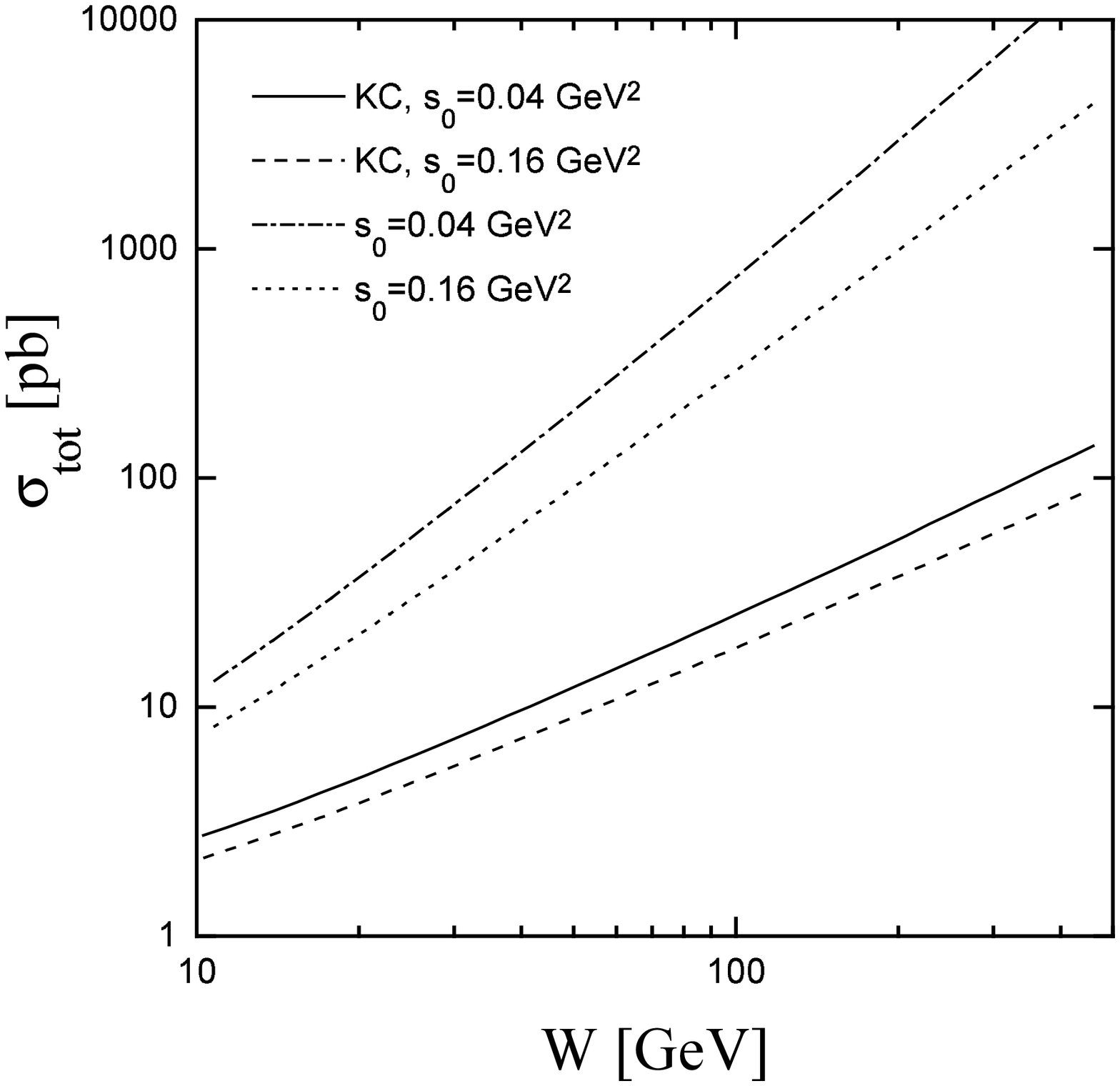}}
\end{center}
\caption{}
\end{figure}
\newpage
\begin{figure*}[tbh]
\begin{center}
\vspace*{-0.6cm}
\epsfig{file=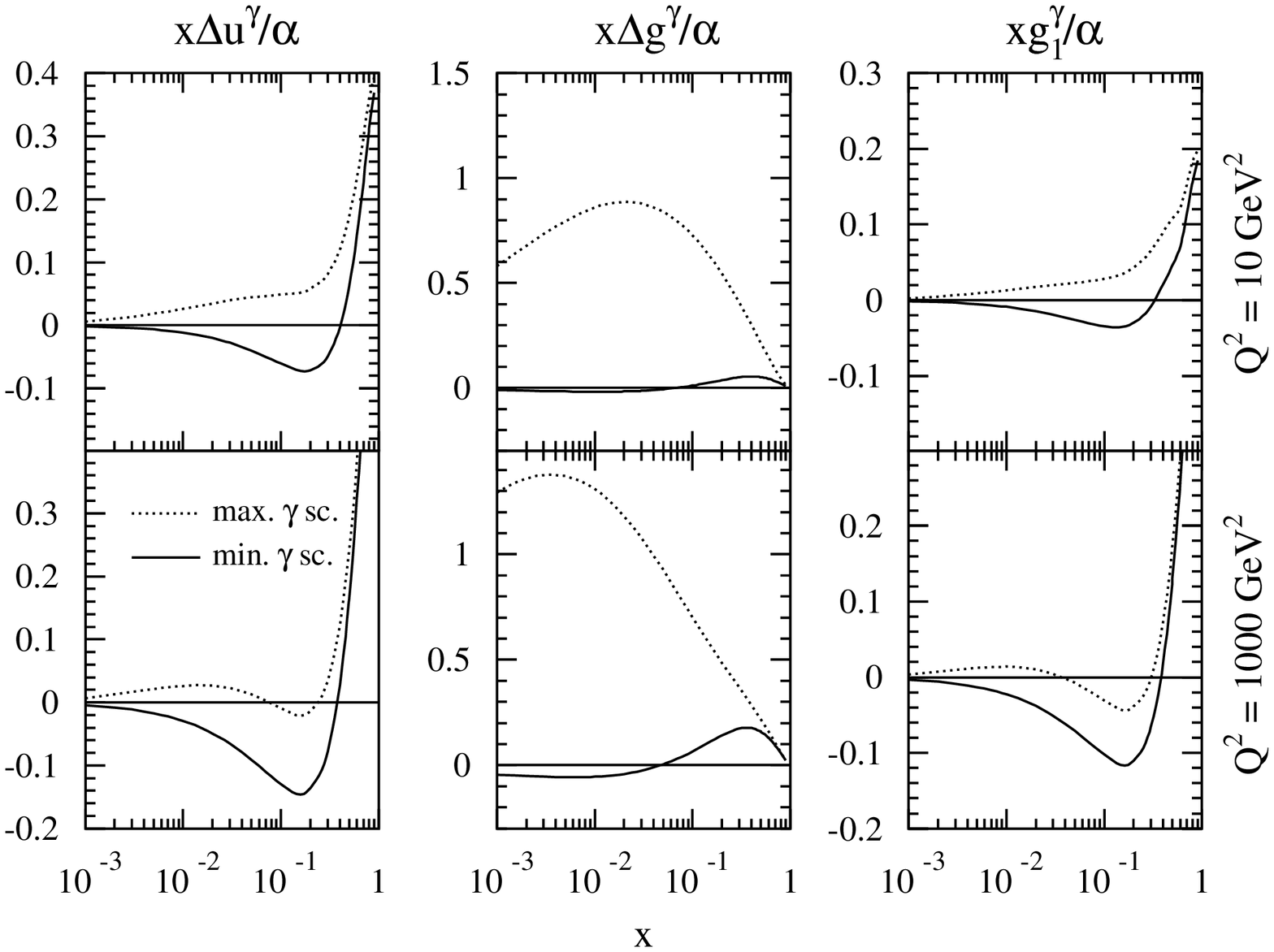,width=13cm}
\end{center}
\caption{}
\end{figure*}
\newpage
\section*{Figure captions}
 \begin{enumerate}
\item Expected event numbers of electron-photon DIS for the 
 backscattered-laser $e\gamma$ mode of a 500 GeV linear collider. It is 
 assumed that 10\% of the $e^+ e^-$ luminosity can be reached in this 
 mode for a rather monochromatic photon beam. (From ref. \cite{PC}.)

\item Energy dependence of the cross-section for the process $\gamma\gamma
\rightarrow J/\psi J/\psi$. The two lower curves correspond to the
calculations based on the BFKL equation with kinematical constraint 
generating the dominant subleading effects 
and the values of $s_0$ equal to $0.04 {\rm \; GeV}^2$
(the continuous line)  and to $0.16 {\rm \; GeV}^2$ (dashed line).
The two upper curves correspond to the BFKL equation in the
leading logarithmic approximation with $s_0=0.04 {\rm \; GeV}^2$
(dash-dotted line) and $s_0=0.16{\rm \;GeV}^2$ (short-dashed line). 
The parameter $s_0$ controlls  the contribution from the infrared 
i.e. confinement region. (From ref. \cite{KMJPSI}.)

\item $x\Delta u^{\gamma}/\alpha$ and $x\Delta g^{\gamma}/\alpha$  
evolved to $Q^2=10$ and $1000\,\mathrm{GeV}^2$ in LO using the two 
extreme inputs (\ref{eq:maxsat}) and (\ref{eq:minsat}).
Also shown is the structure function $g_1^{\gamma}$ in LO. 
(From ref. \cite{GG2}.)
\end{enumerate}       
\end{document}